\documentclass[referee]{raa}
\usepackage{graphicx,times}             
\input{epsf.sty}                      
\input{psfig.sty}                    

\begin{document}
 
\title{A truncated accretion disk in the galactic black hole candidate source H1743-322}

\author{K. Sriram\inst{1}, V. K. Agrawal\inst{2,3}
           \and 
           A.R. Rao\inst{2}
          }
\institute{ Department of Astronomy, Osmania University,
              Hyderabad 500 007, India\\
              {\it astrosriram@yahoo.co.in}\\
\and
            Department of Astronomy and Astrophysics, Tata Institute of Fundamental Research, Mumbai 400005, India \\
\and 
 Indian Space Research Organization, New BEL Road, Bangalore 560 094, India  \\
       }

\abstract{To investigate the geometry of the accretion disk in the source H1743-322, 
we have carried out a detailed X-ray temporal and spectral study using RXTE pointed
 observations. We have selected all data pertaining to the Steep Power Law (SPL)
 state during the 2003 outburst of this source. We find anti-correlated hard X-ray 
lags in three of the observations and the changes in the spectral and timing 
parameters (like the QPO frequency) confirm the idea of a truncated accretion disk 
in this source. Compiling data from similar observations from other sources, 
we find a correlation between the fractional change in the QPO frequency and 
the observed delay. We suggest that these observations indicate a definite size 
scale in the inner accretion disk (the radius of the truncated disk) and we 
explain the observed correlation using various disk parameters like Compton 
cooling time scale, viscous time scale etc..  
}

\authorrunning{Sriram, Agrawal \& Rao }            
\titlerunning{Truncated accretion disk H1743-322 }

\maketitle
\section{Introduction}
\label{sect:intro}
The geometry of the inner accretion disk in Galactic
black hole candidates (GBHCs) is still a matter of debate. All GBHCs
show distinct patterns of temporal and spectral characteristics in different spectral
states, most probably due to the geometric configurations attained
by the disk in each of the spectral states (\cite{esin}). Simultaneous spectral and temporal studies
during the X-ray outburst of GBHCs may give a more clear idea about the 
nature of the disk,
because the sources show state transitions during such episodes. Recently, 
several black hole sources were studied during their outburst period
and all of them show a similar trend in their spectro-temporal evolution
(\cite{kalemci, remillard, belloni, tomsick, rodriguez}). The typical
broad band X-ray spectrum of a GBHC  shows two primary components, a
soft component (from the disk) and a hard component from the Comptonization
process (\cite{shapiro76, sunyaev80}) or from the base of the jet
(\cite{markoff}). \\

One of the important spectral states in GBHCs is the  Very High State
(VHS) or the Steep Power-law State (SPL) (\cite{mcclintock}). Many of the
black hole systems become bright during this state and show the
`C' type  Quasi periodic Oscillations (QPOs with high coherence and
accompanied with a flat top noise) and occasionally the  high frequency
QPOs in their Power Density Spectrum (PDS) (\cite{remillard}). Jets are
the most common phenomenon in GBHCs and the study of hardness-intensity
diagrams (HIDs) shows that the jet switch on/off during this state
(\cite{fender}). The observational results suggest that the hard component
in the SPL state arises due to Comptonization of soft photon originating
from a Keplerian disk (\cite{kubota, done})  and it was found that the
disk is truncated, favoring a disk+quasi-spherical cloud geometry
(\cite{zdziarskib}). A detailed justification for assuming the truncated
accretion disk scenario, particularly in the SPL state, is given in Done
et al. (2007). \\

The cross-correlation between soft and hard X-ray emission gives further
insight
to the truncation disk scenario. Three sources viz., Cyg X-3, GRS 1915+105
and XTE J1554-564 showed anti-correlated hard lags between soft and hard
photons in the time scale of a few hundred to a few thousand seconds
(\cite{choudhurya, choudhuryb, sriram}). XTE J1550-564 and GRS 1915+105
(both the sources were in the SPL state) showed model independent
and dependent spectral changes along with  shifts in the QPO centroid
frequencies. The changes in the QPO centroid frequencies were well correlated
with the soft as well as hard X-ray fluxes. Recently, similar kind of time lags were observed
 in Cyg X-2, a neutron star binary system (\cite{Lei08}). The observed anti-correlated hard lag
was attributed to the viscous time scale during which the truncated
disk makes radial inward/outward movement. Since XTE J1550-564 and
H1743-322 show similar spectral and temporal variability characteristics
(\cite{mcclintocka}), we have searched for the anti-correlated hard
lags between soft and hard X-ray photons in the source H1743-322. The
purpose of searching for anti-correlated hard lag in H1743-322 is to strengthen the idea of the truncated disk
scenario in the SPL state. \\

Despite the lack of dynamical confirmation of the mass of the compact
object in H1743-322, it is believed to be a black hole source due
to its spectro-temporal characteristics (\cite{corbel, kalemci,
mcclintock}). H1743-322 was discovered with {\it Ariel} V all sky
monitor in 1977 (\cite{kaluzienski}) and was  precisely localized by
HEAO-I (\cite{doxsey}). The 2003 outburst of the source was observed in
various wavelengths from X-ray by INTEGRAL (\cite{revnivtsev}) and RXTE
(\cite{markwardt}), infrared (\cite{baba}), optical (\cite{steeghs}), to radio
(\cite{rupena}). The source showed relativistic jet emission (\cite{rupenb})
similar to GRS 1915+105 and XTE J1550-564. During the  2003 outburst,
H1743-322 was in SPL and Thermal Dominated (TD)
states for most of the time and it was occasionally in  the Hard State (\cite{mcclintocka}).\\

In this paper, we have selected all the SPL states during the 2003 outburst
(\cite{mcclintocka}) and searched for the anti-correlated hard
lags between soft (2-5 keV) and hard (20-50 keV) X-ray bands. In three
 observations, we have found lags between soft and hard
X-ray emission and for two observations we found a shift in the 
QPO centroid frequency. We have also found pivoting and marginal pivoting 
pattern in the spectra, similar to Cyg X-3, GRS 1915+105 and XTE J1550-564. 
Detailed spectral studies  have been carried out to understand the
disk-corona configuration in the SPL state.\\ \\

\section{Analysis and Results}
\label{sect:Ana-Res}
We have used {\it Rossi X-ray Timing Explorer} (RXTE) pointed
observations to study the temporal and spectral behavior of the
source H1743-322 and have used data from the Proportional Counter Array
(PCA) (\cite{jahoda}) and the High-Energy X-ray Timing Experiment (HEXTE)
(\cite{rothschild}), effectively covering the 2-150 keV energy band.
H1743-322 was in outburst in 2003 and extensive observations were carried
out using the RXTE satellite. Detailed  spectral and temporal studies
have been carried out and all the observations were classified in SPL,
TD (Thermal Dominated) and Hard spectral states (\cite{mcclintocka}). We
have chosen all the SPL states to look for the anti-correlated hard
lags. We have used Standard 2 data to obtain the light curves and
spectra and followed all the procedure for data filtering and background
corrections. For light curves, data were obtained from all the PCUs
which were ON (it was noticed that most of the time PCU0 and PCU2 were
ON). For obtaining spectra PCU2 was chosen since it is the most well
calibrated among all the PCUs. A 0.5\% systematic error is applied to
the PCA spectra. We have taken the data from single bit mode to obtain
the Power Density Spectra (PDS) for the observations where the lag was
observed. To obtain the 20-150 keV spectra, we have used HEXTE cluster A
data and applied all the necessary corrections. We have used HEASOFT 6.2
software to reduce the data and XSPEC 11.3.1 for spectral analysis. All
the errors mentioned through out the paper are of nominal 90\% confidence
level ($\Delta \chi^2$ = 2.7).\\

\begin{figure}[h]
\includegraphics[height=18cm,width=15cm, angle=-90]{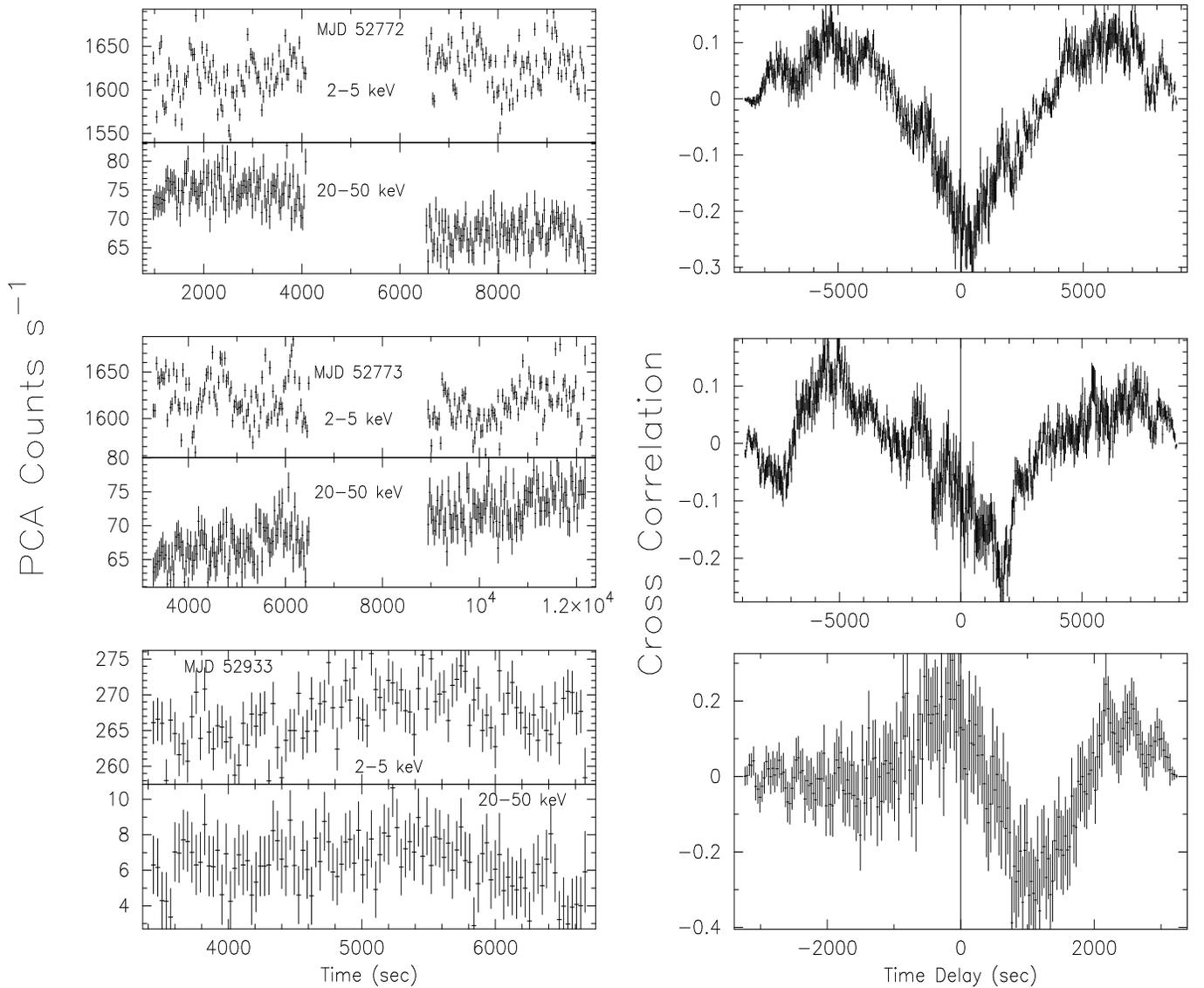}
 \caption{The figure shows the soft and hard X-ray band light curves along with respective cross correlation plots. Left panels: soft and hard X-ray band lightcurves, Right panels: corresponding cross-correlation of soft and hard light curve. The vertical line at zero is drawn for clarity.}    \label{Fig1}
   \end{figure}

For all the SPL state observations, we have extracted the light curves in
two different energy bands. The first energy band spans 2-5 keV (soft)
and the second one spans 20-50 keV (hard). The basic idea of dividing the
light curves in two bands is that the soft band  covers most of the soft
photons originating from the Keplerian disk and the hard band covers most
of the hard photons coming from Comptonization region in the accretion
disk. We have used the {\it crosscor} program provided by Ftools. For
more details see Sriram et al.(2007). Out of 170 observations, 90
were in SPL state and anti-correlated hard lags were found in three
observations. The observed light curves in the two energy  bands are
shown in Fig~\ref{Fig1}, along with the cross-correlation plots. The
last observation is quite soft (with the 20 - 50 keV count rates of only
about 6 s$^{-1}$ compared to the 60 - 80  s$^{-1}$  in the other two
observations) and we note that there were no type `C' QPOs during this
observation. This particular observation is found to be in Hard SPL state (\cite{mcclintocka}).
We have used the method given in \cite{sriram} to measure the
lags and the errors in them and these are give in  Table~\ref{delay}. The
detected lags range from few 100 s to 1000 s, similar to what was
found in XTE J1550-564 and GRS 1915+105.

{\begin{table}{}
\caption{Details of the observed anti-correlated hard lags.}
\label{delay}
\centering
\begin{tabular}{c c c c}
\hline
obsid&delay (sec)&bin (sec) &Correlation Coefficient\\
\hline

$80146-01-36-00$&$442.16\pm37.5$ & 32 & $-0.30\pm0.04$ \\
$80146-01-37-00$&$1590.65\pm57.5$ & 32 & $-0.33\pm0.04$\\
$80137-01-22-00$ & $1129.74\pm42.0$ & 32 & $-0.39\pm0.12$\\
\end{tabular}
\end{table}}

To study the variation of the source properties during the periods
of the observed lags, we have made a detailed analysis of the initial
(part A) and final part (Part B) of the observations, lasting for 300
s each. Examination of the model independent spectra showed that there
were minute changes indicating pivoting and marginal pivoting. Similar
kind of pivoting features were observed in Cyg X-3, XTE J1550-564
and GRS 1915+105 (\cite{choudhurya, choudhuryb, sriram}). In the
first and the second observations, there is a sharp
pivoting around $\sim$6.0 keV whereas no pivoting feature is observed
in the third observation. It should be noted that during all
the observations of the lags, the sources were  in the
SPL/VHS spectral state and the study of different GBHCs suggest that the
thermal Comptonization process is the working mechanism in this state
(\cite{done, kubota, zdziarskic}).\\

We have carried out a detailed spectral analysis to uncover and constrain
the spectral parameters responsible for the anti-correlated hard lags. We
have obtained the spectra covering 2.5-150 keV from the initial and final
part of the light curves and fit the data using the multi-component model
{\it diskbb+thcomp+Power-law}. The {\it diskbb model} (\cite{makishima})
takes care of the soft part of the spectrum, {\it thcomp} (thermal
Comptonization model) (\cite{zdziarskia}) handles the hard spectrum
produced by this process and the Power-law model represents the high-energy
non-thermal photons. Edges and a Gaussian line near 6.4 keV were used
whenever the residuals indicated their presence. Four components were
frozen: hydrogen column density $N_{H}$= 2.2 $\times$ $10^{22}$ $cm^{-2}$,
Gaussian line energy E=6.4 keV and  line width=0.2 keV and the Power-law index
$\Gamma$=2.2. We have verified that keeping these parameters free does
not change the  results appreciably.  The derived spectral parameters
are shown in Table ~\ref{tab2}.


In the first and the second
observations the disk temperature is found to be very low  - 
 $kT_{in}$$\sim$0.30 keV (see Table ~\ref{tab2}). There is a notable change in the electron
temperature in the second observation. The third observation
does not require the {\it thcomp} model, particularly because of the low count
rates in the hard X-ray band. We have used diskbb+Power-law model for this
observation and the Power-law index has steepened from $\Gamma\sim$2.13
to $\Gamma\sim$2.37 whereas the difference in the disk temperature is
less contrasting.

To know which parameters are responsible for the model independent
pivoting, we have fitted the respective spectra of an observation
simultaneously. We have followed the same procedure as used      in our
earlier work (\cite{sriram}). 
 We found that in the first and the second observations, 
the normalizations of the  disk and the thcomp components
have changed during the two different parts of the respective observations
 and it  seems that this change is responsible for
the observed lag and pivoting (see Table~\ref{tab2}). We have not
carried out this method for the third observation because of the 
low count rates at high energies.

We have calculated the unabsorbed disk and thcomp fluxes (see Table
~\ref{tab2}) for the three observations. In the first observation,
the disk and thcomp fluxes are anti-correlated whereas in the second observation
a similar anti-correlation is noticed if we include the Power-law flux too. In
the last observation, the disk and Power-law fluxes are anti-correlated
(see Table ~\ref{tab2}). Similar kind of results were observed in case
of XTE J1550-564 and GRS 1915+105, giving more credence to the truncated
accretion disk geometry in the SPL spectral in these three black hole
sources.\\

\subsection{Quasi-Periodic Oscillations} 
\label{sect:qpo}
\begin{figure}[h] 
\includegraphics[height=10cm,width=10cm,angle=-90]{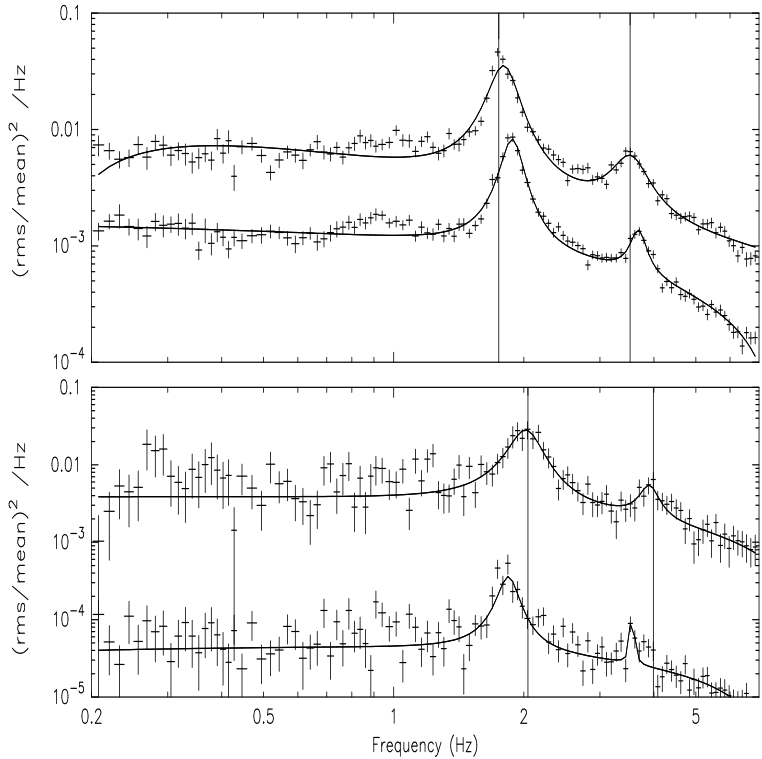}
      \caption{The figure shows the change in the centroid frequency in the initial and final part (part A and B) of the light curve  ObsID 80146-01-36-00 (top panel) and ObsID 80146-01-37-00 (bottom panel). The lines are drawn to mark the shift and part B data is shifted vertically for clarity. }
         \label{Fig2}
   \end{figure}

H1743-322 shows all kinds of QPOs ranging from LFQPOs (Low Frequency
Quasi Periodic Oscillation) to HFQPOS (\cite{homan, remillarda,
mcclintocka}). The origin of HFQPOs is unknown and the production of
these features may be due to some resonance mechanisms (\cite{abramowicz}),
whereas the LFQPOs generally mimics the Comptonization region in the
accretion disk (\cite{chakrabarti,titarchuk}). During the anti-correlated
hard lag, the QPO  centroid frequency changes in GRS 1915+105 and XTE
J1550-564 (Choudhury et al. 2005, Sriram et al. 2007). We have used
single bit mode data (SB\_125us\_8-13\_1s) to obtain the respective PDS
(Power Density Spectrum) of the initial and final parts of the observations in
which lags were detected. In two observations QPO features were  observed
(see Table ~\ref{tab2}) and they are modeled by a Lorentzian function along with
a power law to the continuum. We found that the QPO  centroid frequency
is shifted to either towards low or high frequency (see Fig.~\ref{Fig2}). 
The last observation (ObsID 80137-01-22-00) does
not have a QPO in its PDS. The relative shift of the QPO centroid
frequency between two parts of the observations suggests a change
in the geometrical/physical aspect of the Comptonizing region during the
detected lags.\\

\setcounter{table}{1}
{\begin{table}
\begin{minipage}[t]{\columnwidth}
\caption{Details of the spectral and temporal parameters in individual part of the respective observations. A and B correspond to the initial and final parts of the observation.} 
\label{tab2}
\centering
\renewcommand{\footnoterule}{}
\begin{tabular}{ccccccc}
\hline
parameters&\multicolumn{2}{c}{80146-01-36-00}&\multicolumn{2}{c}{80146-01-37-00}&\multicolumn{2}{c}{80137-01-22-00}\\
\hline
&A&B&A&B&A&B\\
\hline
$kT_{in}\footnote{Disk Temperature using diskbb model}$&$ 0.32\pm0.01$&$0.28\pm0.01 $&$0.28\pm0.01$&$0.29\pm0.01$&$0.66\pm0.02$& $0.63\pm0.02$ \\
$\Gamma_{th}\footnote{thcomp index}$&$ 2.07\pm0.01$&$2.05\pm0.01$&$2.12\pm0.01$&$2.10\pm0.01$\\
$kT_{e}\footnote{Electron Temperature}$&$8.23\pm0.33 $&$6.94\pm0.28$&$9.44\pm0.91$&$8.95\pm0.90$\\
$N_{th}\footnote{thcomp normalization}$&$ 0.74\pm0.01$&$0.75\pm0.26$&$1.04\pm0.20$&$0.85\pm0.24$\\
$\Gamma_{Pl}\footnote{Powerlaw index}$&-&-&-&-&$2.13\pm0.09$&$2.37\pm0.09$\\
$N_{Pl}\footnote{Powerlaw normalization}$&-&-&-&-&$0.31\pm0.01 $&$0.57\pm0.12$\\
$\chi^{2}$/dof&117/80 & 104/80&83/82&114/82&72/84&76.28/84\\
disk flux\footnote{The flux unit for all the models is $10^{-9}$ergs $cm^{-2} s^{-1}$}&61.40& 120.56 & 119.01 & 84.26&4.32&3.96\\
thcomp flux& 10.11& 8.65 & 12.67 & 11.30&-&-\\
Powerlaw flux &10.54&15.32&8.98&13.94&3.96&5.60\\
\hline
Simultaneous fit\\
$N_{th}$&$0.84\pm0.04$&$0.47\pm0.05$&$1.07\pm0.22$&$0.94\pm0.22$&&\\
$N_{bb}/1000$&$ 504\pm31$&$444\pm27$ &$917\pm176$&$587\pm108$&&\\
$kT_{in}$&$0.30$(fix)&$0.30$(fix)&0.28(fix)&0.28(fix)&&\\
$\Delta$$N_{th}$/$N_{th}$(\%)&44&-&12&-&&\\
$\Delta$$N_{bb}$/$N_{bb}$(\%)&12.0&-&36&&&\\
\hline 
Delay (sec)& $442.16\pm37.5$&-&$1590.65\pm57.5$ &-&$1129\pm42.0$&-\\
$\nu$ (Hz)\footnote{QPO centroid frequency}&$ 1.78\pm0.02$&$ 1.87\pm0.02$ &$ 2.02\pm0.02$ &$ 1.86\pm0.02$\\
$\Delta$$\nu$/$\nu$ \%& $5.00$&---&$-7.92$&--\\
\hline
\hline
\end{tabular}
\end{minipage}
\end{table}}

\section{Discussion and Conclusions}
\label{sect:Dis}
The detailed study of the sources GRS 1915+105 and XTE J1550-564  shows
that during the observed lags, the spectral as well as the temporal
properties significantly change, favoring a truncated disk scenario. 
Both the sources show a pivoting feature and changes in the spectral parameters 
during the lags (\cite{choudhuryb, sriram}). The lags obtained between soft and hard flux in SPL/VHS state clearly
suggest that the accretion disk structure changes, may be because of
relative changes in the mass accretion rate. The model independent spectra
show that during the lag, the spectra change and show a pivoting pattern.
The pivoting in the spectra directly gives an
idea of change in the flux in softer and harder part of the spectra. The
detected lags most likely indicate the viscous time scale during which 
the disk front
moves inward/outward depending on the local mass accretion rate and hence the 
soft and hard emission region changes their respective temporal and spectral properties. 

From the spectral analysis of the first observation, we found that the disk temperature 
is not changing
but the electron temperature of the Compton cloud changes ($\delta
kT_{e}\sim$2 keV). In the last observation
the Power-law index changes by $\delta\Gamma\sim$0.2, maintaining
the disk temperature around 0.63 keV (see Table ~\ref{tab2}). 
Perhaps the most important and main support for anti-correlated
lag comes from the observed soft and hard X-ray flux values. It can be seen from
Table ~\ref{tab2} that as the soft flux increases, the correspond
hard flux decreases,  and vice-verse. It suggests that the flux is the
important parameter which changes during the lag. In the first two observations
(see Table ~\ref{tab2}), the QPO  centroid frequency is shifted during the
lags, indicating that the size of the Compton cloud is altering. The soft
flux is well correlated with the shift in the centroid frequency suggesting
changes in the soft and hard emitting region in the accretion disk. \\

\begin{figure}[h] 
\includegraphics[height=15cm,width=10cm,angle=-90]{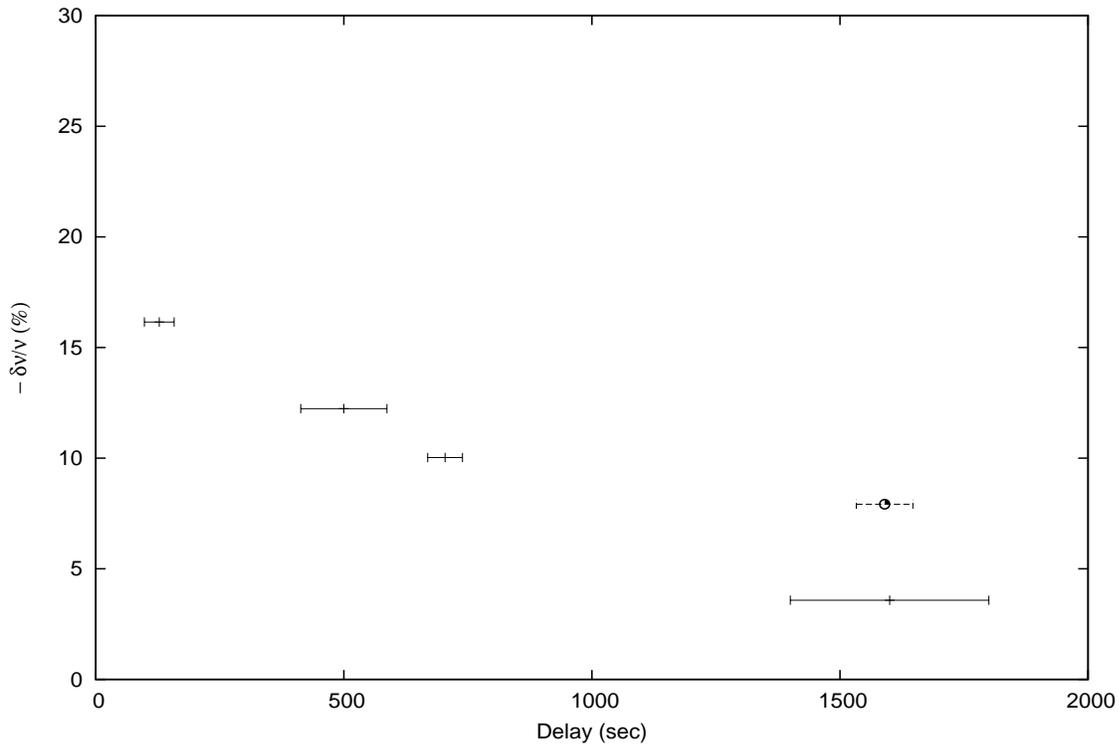}
      \caption{The figure shows a relation between the observed delay and the fractional percentile change (when QPO centroid frequency is decreasing) in the QPO frequency. The circle represents the data point of the second observation of the source H1743-322 and the remaining data points belong to the source GRS 1915+105.}
         \label{Fig3}
   \end{figure}

\begin{figure}[h] 
\includegraphics[height=15cm,width=10cm,angle=-90]{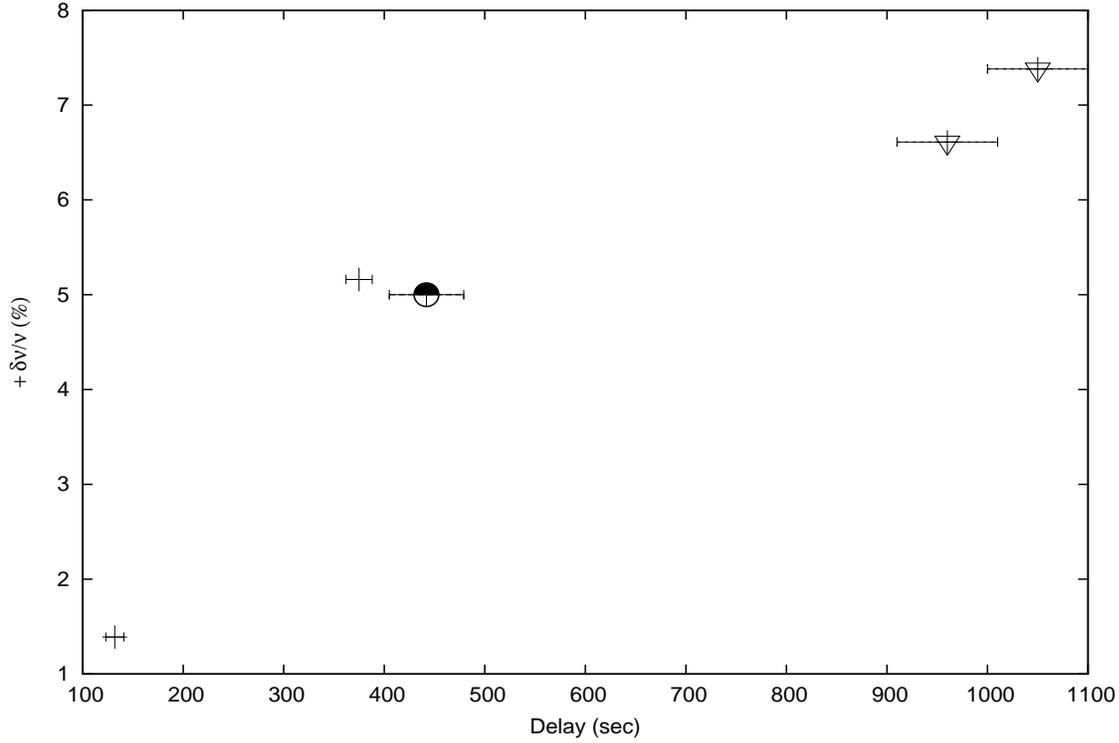}
      \caption{The figure shows a relation between the observed delay and the fractional percentile change (when QPO centroid frequency is increasing). The circle represents the data
point of the first observation of the source H1743-322, inverted triangle represents the data points of GRS 1915+105 and the remaining points are of XTE J1550-564.}
         \label{Fig4}
   \end{figure}

We have also investigated the dependence of the 
change in the QPO frequencies with the delays. In Figure \ref{Fig3} we
have plotted the fractional change (in percentage) in the QPO frequency 
(absolute values) against the observed delays,
for cases where the QPO frequencies were found to be decreasing.
In Figure \ref{Fig4}, a similar plot is shown for cases where the QPO frequencies were found to be increasing.
Apart from the source H1743-322, we have used data from Choudhury et al. (2005)
 for GRS 1915+105 and from Sriram et al. (2007) for the source XTE J1550-564.
 The fractional QPO frequency change (when QPO centroid frequency is decreasing) in Figure
 3 indicates that the disk radius is increasing (moving outwards), whereas in Figure 4,
 fractional QPO frequency change (when QPO centroid frequency is increasing) indicates
 that the disk radius is decreasing (moving inwards). \\

We attempt to give a qualitative description for these variations, based on
some general considerations of a truncated accretion disk.
 First 
we take the case when
the fractional change in QPO frequency is increasing as the delay period is increasing (Figure 4).
We assume that initially the disk is truncated at a large radius and then moves inward 
causing the soft flux to increase. 
The increase in the soft flux causes the hot corona to cool. If the observed QPO is linked to the size of the Compton cloud, 
an increase in the QPO frequency is expected.

The effective delay between two parts of the observation where positive change in the QPO frequency and pivoting in the spectra is observed can be given by

\begin{equation}
t_{delay} = t_{viscous}+|t_{eff}|
\end{equation}
where $t_{eff}$ is effective cooling or heating time scale (will have positive value for cooling and negative value for heating). This time scale is given by 
\begin{equation}
t_{eff}=t_{heat} \times t_{cool}/(t_{heat}-{tcool})
\end{equation}

$t_{heat}$ is the heating time scale given by,

\begin{equation}
t_{heat} = 10^{-3} \alpha^{-1} m_{10}^{-1/2} R_7^{3/2}
\end{equation}
where $\alpha$ is the dimensionless viscosity parameter, m$_{10}$ is mass of black hole 
expressed in terms of m$_{10}$=M/10 M$_{\odot}$, R$_{7}$ is size of the disk in terms of R/(10$^{7}$ cm) (see \cite{frank02}).\\ 
$t_{cool}$ is the  cooling time scale given by,
\begin{equation}
t_{cool} = \frac {N_e k T}{\eta L_{seed}}
\end{equation}
where $N_e$ total number of electrons in Compton cloud, $\eta$ is energy gain factor during Compton scattering and $L_{seed}$ is seed photon luminosity.
Assuming that the seed photons are  supplied by the standard thin accretion disk, the cooling time scale becomes
\begin{equation}
t_{cool} = 10^{-6} \times R_7^3 \dot M_{17}^{-1} m_{10}^{-1} T_8 
\end{equation}
where, $\dot M_{17}$ is mass accretion rate terms of $\dot M$/(10$^{17}$ g s$^{-1}$), T$_{8}$ is electron temperature in terms of T/(10$^{8} K$).\\

For the present case (where QPO frequency is increasing and truncation radius is
decreasing), the cooling is predominant. 
In Figure 5 we have plotted the effective cooling time scale as a function of radius. 
The delay is basically determined by the viscous time scale and hence will be 
large for a large truncation radius. Hence if delay is high,  then a small decrease 
in radius results in a large decrease in the cooling time scale causing the Compton cloud to shrink faster. This will give rise to a larger relative change in QPO frequency when the disk is truncated at a larger radius.


\begin{figure}[h] 
\includegraphics[height=15cm,width=10cm,angle=-90]{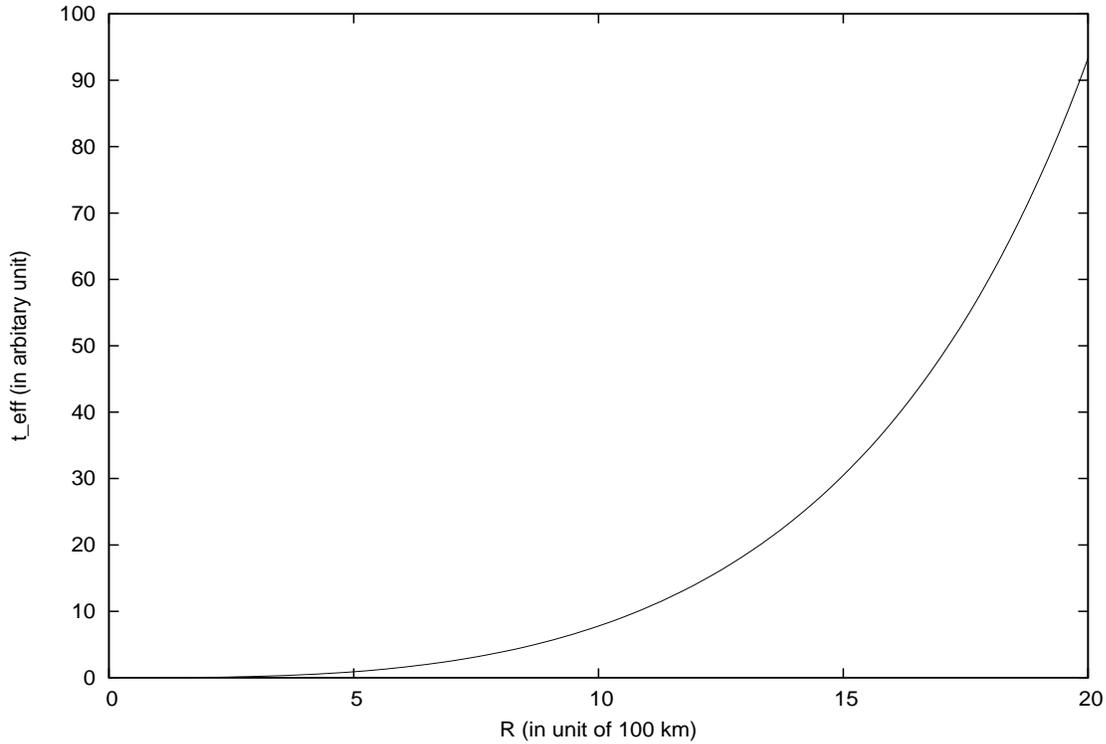}
      \caption{The figure shows the effective cooling/heating  rate,  t$_{eff}$, for cooling (see text).}
         \label{Fig5}
   \end{figure}

Now consider the other case where the negative change (decrease in QPO frequency) 
decreases with the delay (Figure 3). Assume that initially the disk is truncated
 at a small radius and moves outward during the two different parts of an observation,
 causing  the seed photon supply to corona to decrease. Hence this will result in 
an effective heating of the corona and a decrease in QPO frequency is expected.
 We plot the effective heating time scale as a function of truncation radius in Figure 6. In this figure we have taken modulus of $t_{eff}$ to make it positive. When truncation radius (or delay) is small, a minute change in radius causes a large decrease in effective heating time scale causing the corona to expand rapidly which in turn results in a large fractional decrease in QPO frequency. This scenario is evident in Figure 3.


\begin{figure}[h] 
\includegraphics[height=15cm,width=10cm,angle=-90]{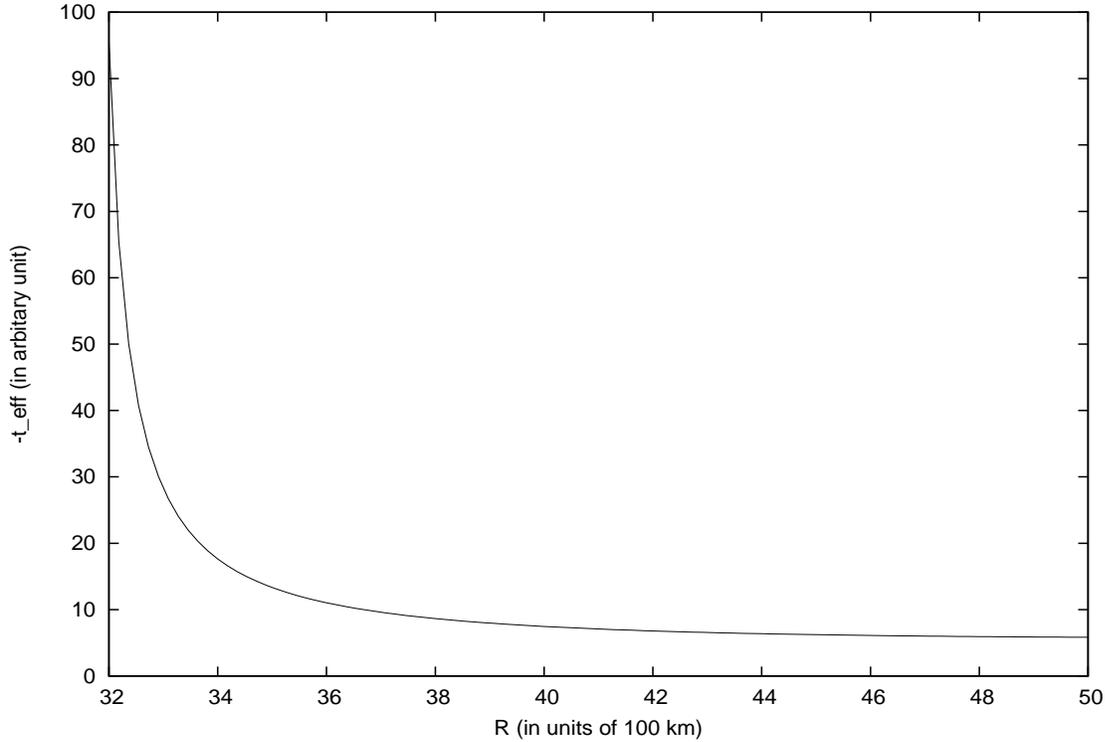}
      \caption{The figure shows the modulus of t$_{eff}$ for heating (see text). }
         \label{Fig6}
   \end{figure}


To obtain the exact value of the truncation radius from the X-ray spectral fitting
is quite difficult because we need to take into account the spectral hardening
due to scattering, relativistic effects and other physical processes occurring
very near to the black hole. But, the inner edge of the truncated accretion disk 
can act as a nozzle to launch the jet and the observed strong correlation
of the X-ray and radio flux in Cyg X-3 and GRS 1915+105 (Choudhury et al. 2003),
 which are
believed to be in the SPL state most of the time, may indicate the very
significant role played by the truncated accretion disk.

In conclusion, the obtained results bias toward a truncated disk scenario favoring a
{\it disk+sphere} geometry. Overall the temporal and the spectral 
observations suggest that during the detected lag, the disk and Compton cloud emission property changes inversely and in this time span the disk is readjusted. The obtained lags are not state transition time scales but the scenario can be viewed as mini state transition during which the properties of the soft and hard emitting region changes.
\begin{acknowledgements}
     This research has made use of data obtained through HEASARC Online Service, provided by the NASA/GSFC, in support of NASA High Energy Astrophysics Programs. K.S is supported by UGC through RFSMS scheme and thankful to TIFR for providing the facility to carry out the work. 
\end{acknowledgements}


\label{lastpage}
\end{document}